\documentclass[manuscript]{aastex}

\slugcomment{To be submitted to the Astrophysical  Journal Letters.}

\shorttitle{Precursor of sunspot penumbral formation}
\shortauthors{Shimizu, Ichimoto, and Suematsu}

\begin{document}


\title{Precursor of Sunspot Penumbral Formation discovered with Hinode SOT Observations}


%
%
%

\author{Toshifumi Shimizu\altaffilmark{1}, 
Kiyoshi Ichimoto\altaffilmark{2}, and Yoshinori Suematsu\altaffilmark{3}}
\altaffiltext{1}{Institute of Space and Astronautical Science,
Japan Aerospace Exploration Agency, 
3-1-1 Yoshinodai, Chuo, Sagamihara, Kanagawa 252-5210, Japan.
Email: shimizu.toshifumi@isas.jaxa.jp}
\altaffiltext{2}{Kwasan and Hida Observatories, Kyoto University,
 Kamitakara-cho, Takayama, Gifu 506-1314, Japan.}
\altaffiltext{3}{National Astronomical Observatory of Japan, 
  Mitaka, Tokyo 181-8588, Japan.}




\begin{abstract}

We present observations of a precursory signature that would be helpful
for understanding the formation process of sunspot penumbrae. 
The {\em Hinode} Solar Optical Telescope successfully captured 
the entire evolution of a sunspot from the pore to a large well-developed 
sunspot with penumbra 
in an emerging flux region appeared in NOAA Active Region 11039. 
We found an annular zone (width 3"-5") surrounding 
the umbra (pore) in Ca {\sc II H} images before the penumbra is
formed around the umbra. The penumbra was developed 
as if to fill the annular zone.  The annular zone
shows weak magnetogram signals, meaning less magnetic flux or 
highly inclined fields there. Pre-existing ambient magnetic field islands
were moved to be distributed at the outer edge of the annular zone and
did not come into the zone. There is no strong systematic flow patterns in
the zone, but we occasionally observed small magnetic flux patches
streaming out. The observations indicate that the annular zone
is different from sunspot moat flow region and that it represents 
the structure in the chromosphere. We conclude that the annular
zone reflects the formation of a magnetic canopy overlying the region surrounding
the umbra at the chromospheric level, much before the formation
of the penumbra at the photospheric level. The magnetic field structure
in the chromosphere needs to be considered in the formation process
of the penumbrae.

\end{abstract}





\section{Introduction}

Sunspots are dark patches on the solar surface and the most
readily visible manifestations of magnetic flux concentrations
 \citep{sol03, tho08}. 
 A well-developed sunspot typically consists of a dark central
region called the umbra which is surrounded by a less dark, 
annular region called the penumbra. 
Sunspots appear with successive series of magnetic flux emergence,
in which magnetic flux rises through the convection zone to the solar
surface and penetrates into the upper atmosphere.  Pores, which are
essentially small sunspots without penumbra or naked umbra, are
first formed at the both edges of emerging flux regions. Pores are
developed to large sunspot with penumbra through the coalescence 
of pores and smaller magnetic flux tubes into a single, growing pore.
When the pore has grown to sufficient total magnetic flux ($1 - 1.5 \times
10^{20}$ Mx), it forms a penumbra in sectors \citep{lek98}.  The formation
of a penumbra is a sudden event, generally within 20-30 minutes, 
and the Evershed flows are observed without delay after the penumbral
formation \citep{lek98, yan03}. Recently \citet{sch10} showed that
the size of the umbral area is unchanged during the growth of 
the penumbra in about 4 hrs, and concluded that the umbra has 
reached an upper size limit (about 4Mm in diameter) and that any newly 
emerging magnetic flux that joins the spot is linked to the process of 
penumbral formation.

However, the formation process of sunspot penumbra is difficult to 
catch observationally, especially with high spatial resolution. Because of
this, we poorly understand the formation process and have not yet
answered fundamental questions, such as why do sunspots 
have penumbrae? and what causes their rapid formation?
Here we present an unique data set from high-cadence filtergraph 
observations of the {\em Hinode} Solar Optical Telescope (SOT), which
successfully captured the entire evolution of a leading sunspot
from the pore to a well-developed sunspot with penumbra (section \ref{sec: obs}). 
We discover 
{\bf an annular}
zone surrounding the umbra 
in Ca {\sc II H} before the penumbra is formed around 
the umbra (section \ref{sec: results}), and discuss what 
is the 
{\bf annular}
structure in section \ref{sec: discussion}.

\section{Observations and data analysis}
\label{sec: obs}

The observations were carried out by SOT \citep{tsu08, sue08, shi08, ich08}  
onboard {\em Hinode} \citep{kos07}. The SOT continuously monitored
NOAA Active Region 11039 from December 29, 2009 to January 2, 2010
with some short interruptions for XRT (X-Ray Telescope) synoptic 
observations. In the SOT field of view, a large emerging flux region 
appeared from  December 30 to 31 and we completely captured
the overall evolution from the birth of pore to the development of 
the large sunspot at the leading area. The region was located at 
S28 W07 on December 30 and at S27 W22 on December 31.
Broadband Filter Imager took Ca {\sc II H} (3968\AA\ , band width: 3\AA\ )
images every 3 minutes and G-band (4305\AA\ , band
width: 8\AA\ ) images every 2 hours with $2 \times 2$ pixel 
summation (0.10896"). Narrowband Filter Imager acquired
longitudinal magnetograms (0.160"/pixel) in Na {\sc I D} (5896\AA\ )
every 3 minutes.
The spectral bandwidth of the Lyot filter is 90m\AA\ . 
The magnetograms were derived 
with the observable named MG4 V/I (Obs ID 85), in which
{\bf a pair of I+V and I-V was measured at $+140$m\AA\ off position 
four times and  $V/I$ was calculated on board after accumulating 
the 4 pairs for better S/N.}
No spectro-polarimter (SP) data was available
in the period before the penumbral formation. 


The time series of SOT images were aligned in respect to the solar rotation,
which was followed by the spacecraft attitude pointing with a tracking 
curve (rotation rate is 0.00014805 deg/s) \citep{shi07}. 
A cross correlation method was applied with a large field-of-view (FOV) to obtain
the series of aligned images.   
In addition, G-band images acquired by XRT \citep{gol07, kan08} were used to 
remove small gradual FOV drifts in the time series due to proper motion of the solar 
features and intensity gradient in the field of view \citep{shi08}. 
The magnetogram conversion to the flux density (gauss or Mx cm$^{-2}$) is determined by
comparing the magnetogram to the magnetic flux density map derived from 
{\bf SP data}
taken at 13:44-14:05 UT on December 31 for the region outside
the sunspot. 



\section{Results}
\label{sec: results}

\begin{figure}[htb]
  \epsscale{1}
     \plotone{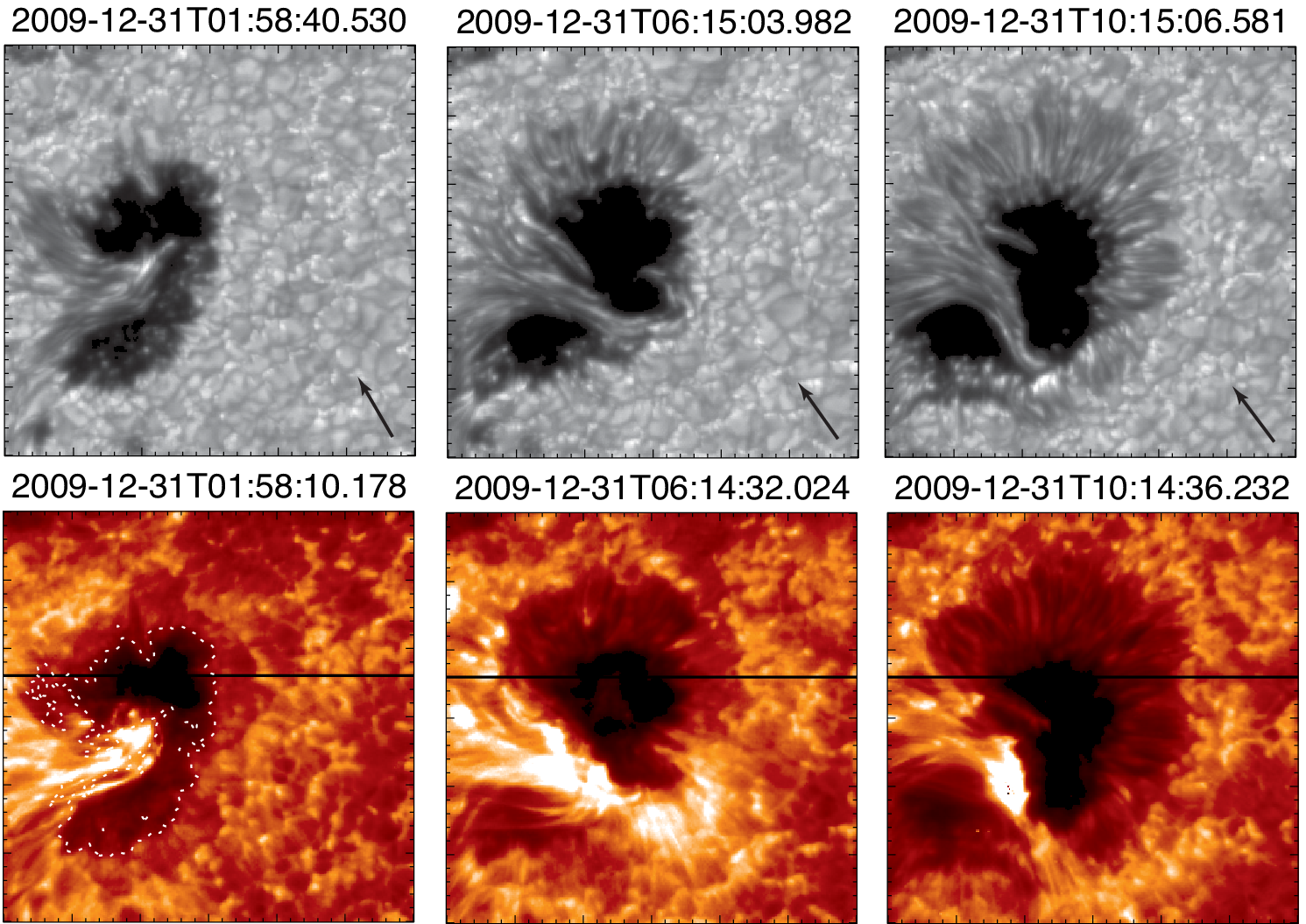}
  \caption{Evolution of the sunspot seen in G-band and Ca {\sc II H} 
    every about 4 hrs. North is up and east is left. The field of view is 
    $32.7" \times 32.7"$ ($300 \times 300$ pixels). The horizontal line
    in Ca {\sc II H} images indicates the position of the slit center for generating the time-slice maps
    in Figure~\ref{fig: ca_tslice} and Figure~\ref{fig: mg_tslice}.
    {\bf The arrows are pointing to disk center and the dotted line is the umbral edge.} 
    }
 \label{fig: gb_ca}
 \end{figure}

\begin{figure}[htb]
  \epsscale{1}
     \plotone{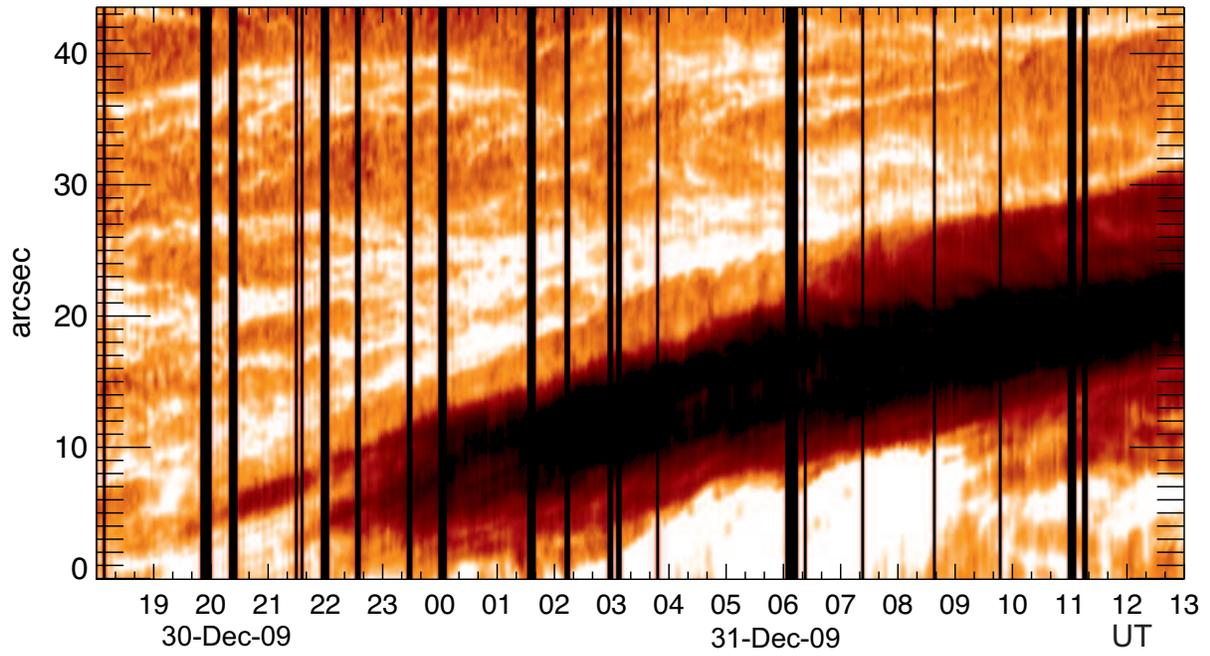}
  \caption{Time-slice of Ca {\sc II H} images for the slit located across the sunspot. 
    The slit center is shown on Ca {\sc II H} images in Figure \ref{fig: gb_ca}. The larger values 
     in the slit position mean toward the west. The black stripes are data gaps. 
    }
 \label{fig: ca_tslice}
 \end{figure}

Figure~\ref{fig: gb_ca} is the temporal evolution of the sunspot seen in 
G-band and Ca {\sc II H} every about 2 hrs. 
Frequent appearance of emerging magnetic patches was observed
at the east side of the spot. The penumbra formed
in sectors; the penumbra was developed at the north side, and then
it formed at the west and south.  Note that remarkable chromospheric 
dynamics associated with the elongated structure
formed in the umbra, is out of scope in this paper and will be
discussed in a separate paper. 
Figure~\ref{fig: ca_tslice} is the time-slice of Ca {\sc II H} images 
for the slit located across the sunspot. The slit center is shown on 
Ca {\sc II H} images in Figure \ref{fig: gb_ca} and the average intensity in
the width of $\pm 7$ pixels from the slit center is given in the time-slice. 
The umbra is drifted toward the west with a speed of 0.22 km s$^{-1}$.
On the slit, the development of the penumbra was 
observed from 6UT to 8 UT on December 31. We can easily notice
{\bf a zone between the umbra and ambient bright features}
before the penumbral formation. 
This 
zone appeared soon after the pore formation
and was seen for about 10 hrs until the penumbral formation. 
The 
zone is annular with the width of 3"-5" in Ca {\sc II H} images.
{\bf In the Ca {\sc II H} image at the left  of Figure~\ref{fig: gb_ca}, 
the zone exists at the outer side of the umbral edge 
(indicated by the dotted contour) on almost all the locations  
excepting the east side of the spot.}
{\bf Its brightness}
is almost similar to that inside the network
cells far from the sunspot. The penumbra was developed as if
to fill the 
annular zone. 
{\bf The zone became very dark after the penumbral formation and
its brightness is $58 - 73$\% of the brightness in
the quiet-Sun network cells. }

\begin{figure}[htb]
  \epsscale{1}
     \plotone{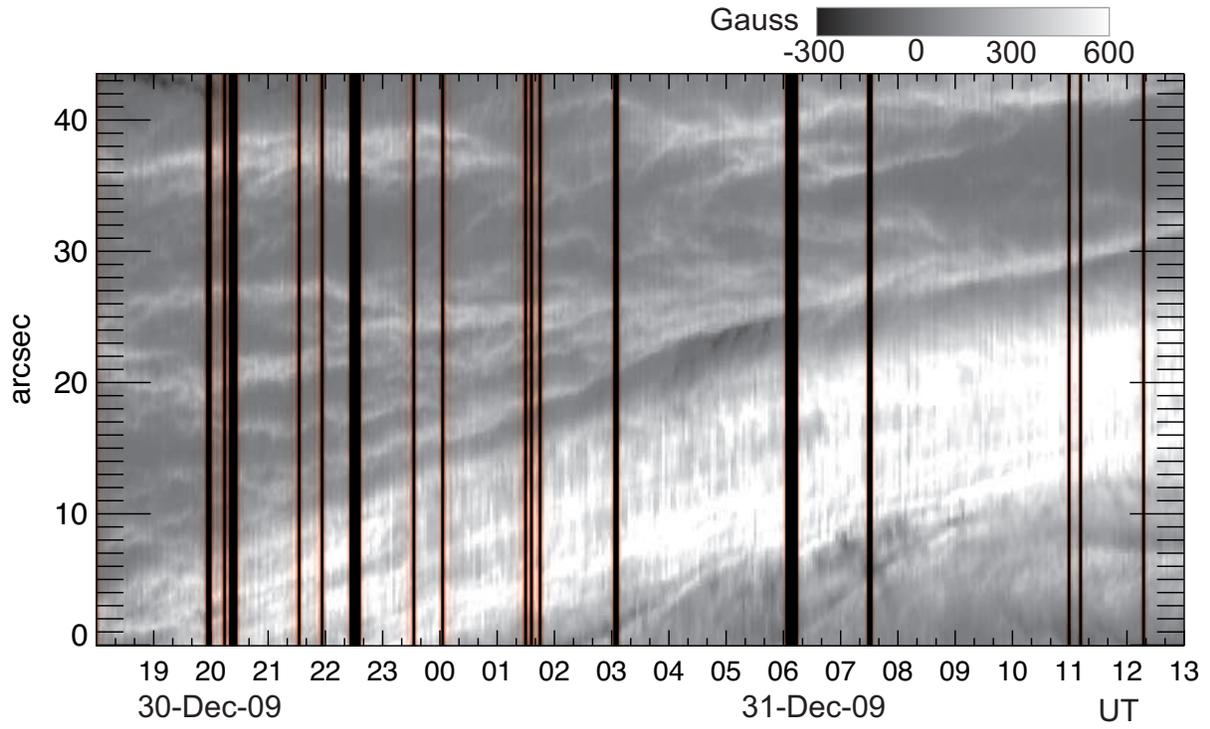}
  \caption{Time-slice of Na {\sc I D} magnetograms. 
    See Figure~\ref{fig: ca_tslice} for details. 
    {\bf Note that the data is clipped as given in the scale bar.}
    }
 \label{fig: mg_tslice}
 \end{figure}

The corresponding time-slice of Na {\sc I D} magnetograms is
shown in Figure~\ref{fig: mg_tslice}. The 
annular zone
shows weak signals; the magnetic flux there is less line-of-sight component
or highly inclined. 
A remarkable behavior is the existence of positive-polarity (the same
polarity with the sunspot)
magnetic flux concentrated at the outer edge of the zone. 
This flux is pre-existing flux patches and may be observed as bright features
in Ca {\sc II H} (Figure~\ref{fig: ca_tslice}).
The sunspot was developed in the region where the dominant magnetic
flux polarity is same as that of the sunspot.  
The pre-existing flux patches that come to the outer edge of the zone
moved to west in response to the motion of the sunspot, and
they did not come into inside the zone. 
The total ambient flux located at the west side of the sunspot is
almost constant ($2.5 \times 10^{20}$ Mx) within $\pm 15$ \% 
in between before and after the penumbral formation.
{\bf Note that we measured the total flux of ambient magnetic patches existing 
in the zone within $\pm 13.6$ arcsec from the position of the slit center 
outside the annular zone or penumbra.}
The total magnetic flux of the sunspot monotonically increased with time and
it was $5 \times 10^{20}$ Mx at 3 UT and $7 \times 10^{20}$ Mx at 6 UT
on December 31. 

\begin{figure}[htb]
  \epsscale{1}
     \plotone{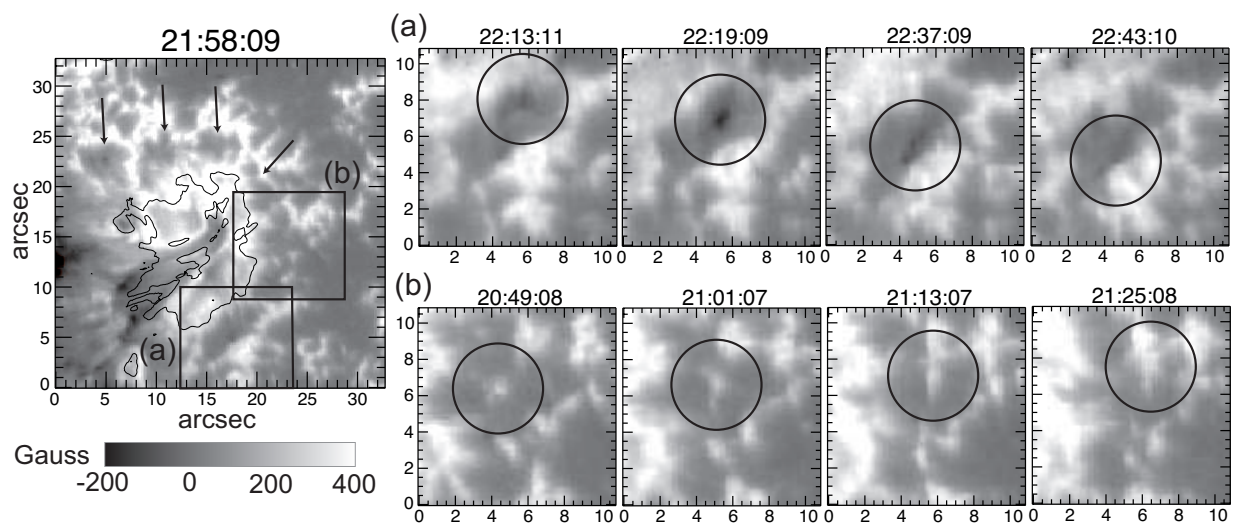}
  \caption{Temporal evolution of Na {\sc I D} magnetograms in two regions 
    located in the annular zone, labeled as a and b. Circules (5 arcsec in diameter)
     indicate the position of an outward moving patch. The field of view is 
    $10.9" \times 10.9"$. North is up and east is left. The solid contour in the large
    field-of-view image gives the outer boundary of the pore determined with 
    a G-band image. 
      }
 \label{fig: pore_mag}
 \end{figure}

Gas flow patterns inside the zone before the penumbral formation were
examined with 3-minutes cadence time series of 
Na {\sc I D} magnetograms (Figure~\ref{fig: pore_mag} and a supplemental movie). 
We observed some flux patches 
that flow out  from the edge of the spot and move outward. 
{\bf Cellular} patterns, 
in which most of flux is distributed at 
{\bf cellular} edge
, were formed in between the spot and the ambient field (arrows in Figure~\ref{fig: pore_mag}). 
The {\bf cellular} patterns 
were gradually evolved to an annular structure. 
There were no strong systematic flow patterns in the annular zone, which is
quite different from the moat region, but 
{\bf a limited number of } 
outward moving patches were observed, as
shown in Figure~\ref{fig: pore_mag} (a) and (b). The speed of these moving 
patches is $1-2$ km s$^{-1}$. In the zone, no magnetic patches were observed to show
inward motion to the umbra. 
{\bf It is noted that the flux patches moving toward the umbra were
frequently observed on the east side of the spot where flux emergence
took place actively\citep{sch10b}.}
After the penumbral formation, 
we observed that small magnetic patches frequently flow out from around
the penumbral edge. 

\section{Discussion and conclusions}
\label{sec: discussion}

Ca {\sc II H} images newly revealed that 
{\bf an} 
annular zone exists
surrounding the pore before the formation of penumbra. 
Here we discuss the physical nature of the annular zone. 

First, the annular zone is the structure different from sunspot moat
region. The moat region is an annular region outside 
the penumbra with a persistent large-scale radial outflow (moat flows) and 
{\bf frequent appearance of}
moving magnetic features \citep{she69, har73}.
Since the penumbra is developed
as if to fill the pre-existing annular zone
{\bf and a systematic moat-flow-like outflow is not observed in the annular zone,} 
we can conclude that 
the annular zone represents the structure different from the moat. 
Second, the annular zone has its origin in
higher atmosphere, i.e.,
chromosphere. This is because the zone can be found in chromospheric
Ca {\sc II H} images but the corresponding signature cannot be 
seen in photospheric G-band images. 

\begin{figure}[htb]
  \epsscale{0.7}
     \plotone{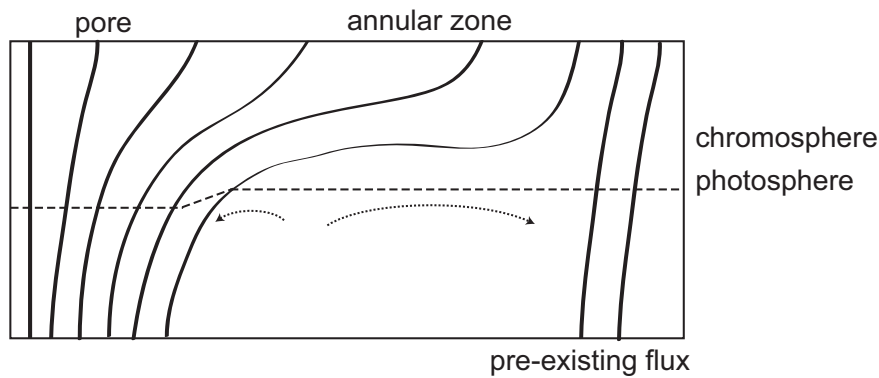}
  \caption{Magnetic field structure before the penumbral formation. The nearly
     horizontal dash line indicates the photospheric ($\tau = 1$) level.  The dotted
     lines with the arrow head are large-scale gas flows  in the subsurface layer.  
    }
 \label{fig: cartoon}
 \end{figure}

A possible interpretation is that the annular zone may reflect 
the formation of a magnetic canopy at the chromospheric level
overlying the region surrounding the umbra, much before the formation
of the penumbra at the photospheric level (Figure~\ref{fig: cartoon}). 
 \citet{hur00} conducted numerical simulations on axisymmetric flux tubes
in a compressible convecting atmosphere and shows how the magnetic
field configuration changes as a function of magnetic flux, i.e., from 
a small pore to a well-developed sunspot. The potential magnetic field
is used in the atmosphere above the photospheric layer. Because of
the pressure balance with surrounding gas that has a decreasing
pressure at higher atmosphere, the flux tube forms a canopy 
configuration in the layer above the photosphere. They showed that 
the size of the canopy structure is almost independent of the total flux 
content of the flux tube, which is in good agreement with the almost constant  
width of the annular zone (3-5") in the entire period from the pore formation 
to the penumbral formation. Because of magnetic pressure of the canopy 
fields, ambient pre-existing magnetic flux elements cannot move into 
inside the annular zone. 

In the penumbrae, the average inclination of the magnetic field 
increases from approximately 40 deg at the umbra-penumbra boundary
to about 70 deg at the outer edge of penumbra 
{\bf \citep{bel03}. }
There is no obvious gradual increase in the field inclination during 
the formation of penumbrae \citep{lek98}. Thus, the appearance
of penumbra would be quite sensitive to the field inclination at 
the edge of the umbra. The magnetic fields at the edge of umbra are 
extended to the chromosphere and corona.
The magnetic field structures created in the chromosphere 
may give influence to the field inclination at the photosphere.
According to recent numerical simulations of sunspot models,
the development of penumbra appears to be quite sensitive to 
the magnetic field structure in the chromosphere.
\citet{rem09} provided the first MHD simulations of
the entire sunspot structure that resolve sunspot
fine structures and their dynamics. The vertical domain size of their
simulations is still small and the top boundary condition is
located only about 700 km above the quiet-Sun $\tau = 1$ 
level. 
With a different boundary condition at the top, the numerical runs 
result in a different penumbral structure of the simulated 
sunspot, including no penumbral formation \citep{rem11b}. 
Rapid penumbral decay is also observed in association with major flares,
suggesting that the penumbral structure at the photospheric
level can be significantly influenced by magnetic field topology 
in the corona \citep{wan04, den05}.

It was observed that 
{\bf a limited number of}
small magnetic flux patches stream out 
at the photosphere below the chromospheric canopy structure. 
{\bf The frequency of the outgoing patches tells that the emergence 
from below the photosphere in the annular zone is not the origin 
for the appearance of the penumbral magnetic features. }
Earlier observations show that motions toward the pore dominate in the 1"-2"
zone around the pore boundary and their average speed is 
0.3-0.5 km s$^{-1}$, while at larger distances the granules 
move away from the pore with the speed much slower than 1 km s$^{-1}$
\citep{kei99, sob99}. Bipolar moving magnetic features were observed
to stream out from a pore \citep{zuc09}. 
Recently, \citet{rem11} made the simulation run to explore the subsurface field and flow structure
around the spot without penumbra and compared it 
with the structure in the run with penumbra. Even in absence of
a penumbra, the simulation shows a large-scale radial outflow
surrounding the sunspots everywhere in the photosphere, 
in addition of a converging flow in the proximity of sunspots. 

Our observations suggest that the canopy structure is formed
in the upper atmosphere, i.e., chromosphere. When unknown conditions
are settled in,  the chromospheric canopy fields may be evolved to the highly 
inclined penumbral fields at the photospheric level. 
{\bf \citet{rez12} found that, at the early stage of the penumbral formation, 
the magnetic area of the umbra extends over the visible limits of 
the penumbra at the photospheric level. There the field strength is fairly 
high and the field has large inclinations. This is consistent with a forming canopy.}
Thus, the observed annular zone
structure can be interpreted as the precursor of the penumbral formation. 
The chromospheric structure surrounding the umbra can depend on
the spatial distribution of pre-existing magnetic flux elements surrounding the umbra and their
magnetic polarity. In the case presented in this paper, the pre-existing
flux is same as the umbra's, and therefore an annular canopy structure is
formed in the area between the flux-concentrated umbra and 
ambient flux patches, as illustrated in Figure~\ref{fig: cartoon}. 
The existence of ambient flux patches may keep the canopy fields from going
down to the photospheric level. 
If the ambient field has opposite sign, an annular canopy structure
may not be well developed around the umbra, because magnetic
reconnection can easily take place in between the umbral field and
ambient field.  Thus, we point out the possibility that different development of chromospheric
structures around the umbrae may lead to different development of
penumbrae, which should be studied observationally with more examples in the future.

We conclude that the 
annular feature in Ca {\sc II H} 
reflects the formation of a magnetic canopy overlying the region
surrounding the umbra at the chromosphere and that the canopy
structure may play an important role in the formation of penumbrae.
Further investigations in the magnetic structures at both the chromosphere
and the photosphere are urgently needed in coming years. 


\acknowledgments
The authors would like to express their thanks to Dr. Y. Katsukawa for
data analysis and Drs. M. Rempel and B. Lites for fruitful comments. 
Hinode is a Japanese mission developed and launched by ISAS/JAXA, with NAOJ
as domestic partner and NASA and STFC (UK) as international partners. It is
operated by these agencies in co-operation with ESA and NSC (Norway). 
This work was partially supported by KAKENHI 23540278.  




\clearpage



\clearpage











\begin{thebibliography}{}

\bibitem[Bellot Rubio et al. (2003)]{bel03} 
  Bellot Rubio, L.R. et al. 2003, \aap, 403, L47
  
\bibitem[Deng et al. (2005)]{den05} 
  Deng, N. et al. 2005, \apj, 623, 1195

\bibitem[Golub et al. (2007)]{gol07} 
 Golub, L. et al., 2007, Solar Physics, 243, 63

\bibitem[Harvey \& Harvey (1973)]{har73} 
 Harvey, K. \& Harvey, J. 1973, Solar Physics, 28, 61

\bibitem[Hurlburt \& Rucklidge (2000)]{hur00} 
  Hurlburt, N.E. \& Rucklidge, A.M. 2000, \mnras, 314, 793
  
\bibitem[Ichimoto et al.(2008)]{ich08} 
  Ichimoto, K. et al. 2008, Solar Physics, 249, 233

\bibitem[Kano et al.(2008)]{kan08} 
  Kano, R. et al. 2008, Solar Physics, 249, 263

\bibitem[Keil et al. (1999)]{kei99} 
  Keil, S.L., Balasubramaniam, K.S., Smaldone, L.A. \& Reger, B. 
  1999, \apj, 510, 422

\bibitem[Kosugi et al. (2007)]{kos07} 
  Kosugi, T. et al. 2007, Sol. Phys., 243, 3


\bibitem[Leka \& Skumanich (1998)]{lek98} 
  Leka, K.D. \& Skumanich, A. 1998, \apj, 507, 454

\bibitem[Rempel et al.(2009)]{rem09} 
  Rempel, M. et al. 2009, Science, 325, 171 
  
\bibitem[Rempel (2011a)]{rem11} 
  Rempel, M. 2011a, \apj, 740, 15 
  
\bibitem[Rempel (2011b)]{rem11b} 
  Rempel, M. 2011b, private communication

\bibitem[Rezaei et al.(2012)]{rez12} 
  Rezaei, R., Bello Gonz\'alez, N. \& Schliehenmaier, R. 2012, \aap, 537, A19 

\bibitem[Schlichenmaier et al.(2010a)]{sch10} 
  Schlichenmaier R., Rezaei, R., Bello Gonz\'alez, N., \& Waldmann, T.A. 2010, \aap, 512, L1

\bibitem[Schlichenmaier et al.(2010b)]{sch10b} 
  Schlichenmaier R., Bello Gonz\'alez, N., Rezaei, R., \& Waldmann, T.A. 2010, Astron. 
  Nachr., 331, 563

\bibitem[Sheeley (1969)]{she69} 
  Sheeley, , N.R.Jr. 1969, Solar Physics, 9, 347

\bibitem[Shimizu et al.(2007)]{shi07} 
  Shimizu, T. et al. 2007, PASJ, 59, S845

\bibitem[Shimizu et al.(2008)]{shi08} 
  Shimizu, T. et al. 2008, Solar Physics, 249, 221

\bibitem[Sobotka et al. (1999)]{sob99} 
  Sobotka, M. et al. 1999, \apj, 511, 436

\bibitem[Solanki (2003)]{sol03} 
  Solanki, S.K. 2003, \aapr, 11, 153

\bibitem[Suematsu et al.(2008)]{sue08} 
  Suematsu, Y. et al. 2008, Solar Physics, 249, 197

\bibitem[Tsuneta et al.(2008)]{tsu08} 
  Tsuneta, S. et al. 2008, Solar Physics, 249, 167

\bibitem[Thomas \& Weiss (2008)]{tho08} 
  Thomas, J.H. \& Weiss, N.O. 2008, Sunspots and Starspots, Cambridge 
  University Press: Cambridge

\bibitem[Wang et al. (2004)]{wan04} 
  Wang, H. et al. 2004, \apj, 601, L195

\bibitem[Yang et al. (2003)]{yan03} 
  Yang, G., Xu, Y., Wang, H., \& Denker, C. 2003, \apj, 597, 1190

\bibitem[Zuccarello et al. (2009)]{zuc09} 
 Zuccarello, F. et al. 2009, \aap, 500, L5


    
\end{thebibliography}
\end{document}